\documentclass[12pt,a4paper]{article}
\usepackage{amsmath,amsthm,amsfonts,amssymb,cite}

\def \al{\alpha}
\def \be{\beta}
\def \ga{\gamma}
\def \dl{\delta}

\def \la{\lambda}
\def \si{\sigma}

\def \Dl{\Delta}

\def \cG{\cal G}
\def \CN{\cal N}

\def \CM{{\cal M}}

\def \CU{{\cal U}}
\def \CV{{\cal V}}

\def \1{^{-1}}
\def \cd{\partial}

\def \dd{{\rm d}}

\def \F{{\rm F}\,}
\def \B{{\rm B}\,}

\def \h{\widehat}

\def \l{\langle}
\def \r{\rangle}
\def \Tr{{\rm Tr}\,}
\def \tr{{\rm tr}\,}
\def \diag{{\rm diag}\,}

\def \d{\dagger}

\begin{document}
\title{
$${}$$
{\bf
A PATH INTEGRATION APPROACH} \\
{\bf TO THE CORRELATORS OF}\\
{\bf XY HEISENBERG MAGNET AND 
}\\{\bf  RANDOM WALKS
}}

\vskip1.5cm

\author{{\bf N. M. BOGOLIUBOV$^*$ and C. MALYSHEV$^\dagger$}\\[1.0cm]
{\small St.-Petersburg Department, Steklov Mathematical Institute,
RAS}\\ [0.0cm]
{\small  Fontanka 27, St.-Petersburg, 191023, RUSSIA} \\                         
[0.5cm]
$^*$ E-mail: {\it bogoliub@pdmi.ras.ru}\\
$^\dagger$ E-mail: {\it malyshev@pdmi.ras.ru} }

\date{}

\maketitle
\vskip1.5cm
\renewcommand{\abstractname}{\centerline{\normalsize
{\bf{Abstract}}}}
\begin{abstract}
\noindent The path integral approach is used for the calculation of the
correlation functions of the $XY$ Heisenberg chain. The obtained
answers for the two-point correlators of the $XX$ magnet are of
the determinantal form and are interpreted in terms of the
generating functions for the random turns vicious walkers.
\end{abstract}

\leftline{\emph{{\bf Keywords:}} Path Integration; XY Heisenberg Magnet; 
Random Walks.}



\thispagestyle{empty}

\newpage

\section{Introduction}\label{bm:sec1}

The problem of enumeration of paths of {\it vicious walkers} on
one-dimensional lattice was formulated by M. Fisher \cite{fish} and
since then continues to attract much attention (see refs. in
\cite{bog, bog1}). The walkers are called `vicious' because they
annihilate each other at the same lattice site, and their
trajectories are thus non-intersecting. Similar problems appear in
the theory of domain walls \cite{fh}, directed percolation
\cite{dom}, self-organized criticality \cite{bak}, and polymer
theory \cite{es}. It has been proposed Ref. \cite{bog} to use the
$XX$ Heisenberg chain to enumerate the paths of the random turns
vicious walkers.

Approach based on path integration was developed in Refs.
\cite{m2, m3} to calculate thermal correlation functions of the
$XY$ Heisenberg magnet. Dependence of the integration variables on
the imaginary time is defined by special quasi-periodicity
conditions. In the present paper, this method is used for the
calculation of the two-point correlation functions of the $XX$ model
and the interpretation of the obtained answer in terms of the
generating functions of the random turns vicious walkers is given.

\section{The problem}\label{bm:sec2}

The Hamiltonian of the periodic $XY$ Heisenberg chain of ``length''
$M$ ($M$ is chosen to be even) in transverse magnetic field $h>0$
is:
\begin{equation}
\begin{array}{rcl}
H&=&H_0\,+\,{\displaystyle\gamma H_1\,-\,h\,S^z}\,,\qquad
{\displaystyle H_0\,\equiv\,-\,\sum\limits^{M}_{n, m=1}
{\Dl}^{(+)}_{n m}\si^+_{n}\si^-_{m}}\,,\\[0.0cm]
H_1&\equiv&\displaystyle{-\,\frac12\sum\limits_{n, m=1}^M
{\Dl}^{(+)}_{n
m}(\sigma^+_n\sigma^+_{m}\,+\,\sigma^-_n\sigma^-_{m})}\,,\quad
\displaystyle{
S^z\,\equiv\,\frac{1}2\sum\limits_{n=1}^M\sigma^z_n}\,.
\end{array}
\label{bm:eq1}
\end{equation}
Here $S^z$ is $z$-component of the total spin operator, and the
entries of the so-called {\it hopping matrix} ${\Dl}^{(s)}_{n m}$
($s=\pm$) are:
\begin{equation} 2{\Dl}^{(s)}_{n
m}\,\equiv\,{\displaystyle\dl_{|n-m|, 1}\,+\,s\dl_{|n-m|, M-1}}\,,
\label{bm:eq2}
\end{equation}
where ${\dl}_{n, l}$ is the Kronecker symbol. The Pauli matrices
$\sigma^\pm_n=(1/2)(\sigma^x_n\pm i\sigma^y_n)$ and $\si^z_n$,
where $n\in {\CM}\equiv \{1,\ldots,M \}$, satisfy the commutation
relations: $\bigl[\si^+_k,\,\si^-_l\bigr] = \dl_{k l}\,\si^z_l$
and $\bigl[\si^z_k,\,\si^\pm_l\bigr] = \pm\,2\,\dl_{k
l}\,\si^\pm_l$. The periodic boundary condition reads:
$\sigma^\al_{n+M}= \sigma^\al_n$, $\forall n$. The Hamiltonian $H$
(\ref{bm:eq1}), taken at $\gamma=0$ (the case of $XX$ magnet),
commutes with $S^z$.

Time-dependent thermal correlation functions are defined as
follows:
\begin{equation}
G_{a b}(m, t)\,\equiv\,Z^{-1}\Tr\bigl(\si^a_{l+m} (0)\,\si^b_l
(t)\,e^{-\be H}\bigr)\,,\quad Z\equiv\Tr\bigl(e^{-\be H}\bigr)\,,
\label{bm:eq3}
\end{equation}
where $\si^b_l (t)\,\equiv\,e^{i t H}\,\si^b_l\,e^{-i t H}$,
$\be=1/T$ is inverse temperature, and $t$ is time. This correlator
may be rewritten in terms of the canonical lattice Fermi fields
$c_i$, $c^\dagger_j$, where $i$, $j\in{\CM}$, by means of the
Jordan-Wigner map:
\[
\si^+_n\,=\,\Bigl(\prod\limits_{j=1}^{n-1}
            \si^z_j\Bigr)\,c_n\,,
\quad \si^-_n\,=\,c_n^\dagger\,\Bigl(\prod\limits_{j=1}^{n-1}
            \si^z_j\Bigr)\,,
\qquad n\in {\CM}\,,
\]
where $\si^z_j = 1\,-\,2 c_j^\dagger c_j$. The periodic conditions
for the spin operators result in the boundary conditions for the
fermions:
\begin{equation}
c_{M+1}\,=\,\bigl(-1\bigr)^{\mathcal{N}}\,c_1\,,\qquad
       c^\dagger_{M+1}\,=\,c^\dagger_1\,\bigl(-1\bigr)^{\mathcal{N}}\,,
\label{bm:eq4}
\end{equation}
where $\mathcal{N}=\sum_{n=1}^M c_n^\dagger c_n$ is the operator of
the total number of particles. In the fermionic representation, $H$
(\ref{bm:eq1}) will take a form $H = H^+ P^++ H^-P^-$, where $P^\pm
= (1/2) ({\mathbb I}\pm (-1)^{\mathcal{N}})$ are projectors
\cite{m2}. The operators $H^s$ are of identical form with $s=\pm$
pointing out a correspondence between these operators and
appropriate specification of the conditions (\ref{bm:eq4}): $c_{M+1}
= -\,s\,c_1$, $c^\dagger_{M+1} = -\,s\,c^\dagger_1$.

Equation (\ref{bm:eq3}) for the $z$-components of spins, for
instance, becomes:
\[
G_{z z}(m, t)\,=\,1\,-\,2\,
Z^{-1}\Tr\bigl(c^\dagger_{l+m}\,c_{l+m}\,e^{-\be H}\bigr)\,
-\,2\,Z^{-1}\Tr\bigl(c^\dagger_l\,c_l\,e^{-\be H}\bigr)
\]
\begin{equation}
 +\,4\,Z^{-1}\Tr\bigl(
c^\dagger_{l+m}\,c_{l+m}\,e^{i t H}\,c^\dagger_l\,c_l\,e^{-(\be+i
t) H}\bigr)\,. \label{bm:eq5}
\end{equation}
To evaluate (\ref{bm:eq5}), it is convenient to consider the
generating functional:
\begin{equation}
{\cG}\,\equiv\,{\cG}(S, T \big|\,\mu, \nu)\,=\,
Z^{-1}\Tr\bigl(e^S\,e^{-\mu H}\,e^T\,e^{-\nu H}\bigr)\,,
\label{bm:eq51}
\end{equation}
where $\mu$, $\nu$ are the complex parameters, $\mu +\nu =\be$. Two
operators, $S \equiv c^\dagger {\h S} c$ and $T \equiv c^\dagger {\h
T} c$, are defined through the matrices ${\h S}\,=$
$\diag\,\bigl\{S_1, S_2, \dots,S_M\bigr\}$, ${\h T}\,=$
$\diag\,\bigl\{T_1, T_2, \dots,T_M\bigr\}$. For instance, the last
term in R.H.S. of (\ref{bm:eq5}) is obtained from (\ref{bm:eq51}) in
the following way:
\[
\frac\cd{\cd S_k}\,\frac\cd{\cd T_l}\,{\cG}(S, T \big|\,\mu,
\nu)\Bigg|
_{\begin{matrix} S_n,\,T_n,\,\forall n &\longrightarrow& 0 \\
             \mu,\,\nu &\longrightarrow&
-i t,\,\be + i t \end{matrix}}\,.
\]
As a result, we express the trace in R.H.S. of (\ref{bm:eq51}) in
the form \cite{m2}:
\begin{equation}
\Tr\bigl(e^S\,e^{-\mu H}\,e^T\,e^{-\nu H}\bigr)\,=\,
   \frac12\,\Bigl({\cG}^+_{\F}\,Z^+_{\F}\,+\,{\cG}^-_{\F}\,Z^-_{\F}
\,+\,{\cG}^+_{\B}\,Z^+_{\B}\,-\,{\cG}^-_{\B}\,Z^-_{\B}\Bigr)\,,
\label{bm:eq6}
\end{equation}
where
\begin{equation}
\begin{array}{rcl}
{\cG}^\pm_{\F}\,Z^\pm_{\F} &=&\Tr\bigl(e^S\,e^{-\mu
H^\pm}\,e^T\,e^{-\nu
H^\pm}\bigr)\,,\\[0.2cm]
{\cG}^\pm_{\B}\,Z^\pm_{\B}&=&\Tr\bigl(e^S\,e^{-\mu
H^\pm}\,e^T\,(-1)^{\CN}\,e^{-\nu H^\pm}\bigr)\,,
\end{array}\label{bm:eq7}
\end{equation}
and $Z^\pm_{\F} = \Tr(e^{-\be H^\pm})$, $Z^\pm_{\B} =
\Tr\bigl((-1)^{\CN}e^{-\be H^\pm}\bigr)$.

\section{The path integral}\label{bm:sec3}

We use the coherent states $| z\big\r\,\equiv\,\exp \bigl(c^\dagger
z\bigr)| 0 \big\r$ and $\big\l z^* |\,\equiv\,\big\l 0 |\exp
\bigl(z^* c\bigr)$ generated from the Fock vacuum $| 0\big\r$, $c_k
| 0\big\r=0$, $\forall k$. We use the short-hand notations for the
$M$-component objects, say, $z^*\equiv(z^*_1,\ldots,z^*_M)$ and
$z\equiv(z_1,\ldots,z_M)$ formed by the independent Grassmann
parameters $z_k,z^*_k$ ($k\in {\CM}$). Besides, $\sum_{k=1}^{M}
c^{\d}_k z_k \equiv c^{\d} z$, $\prod_{k=1}^M {\dd} z_k \equiv {\dd}
z$, etc. Then, we shall represent \cite{m3} the trace of the
operator in $\cG^\pm_{\F}\,Z^\pm_{\F}$ (\ref{bm:eq7}) by means of
the Grassmann integration over ${\dd}z$, ${\dd}z^*$:
\begin{equation}
{\cG}^\pm_{\F}\,Z^\pm_{\F}\,=\, \int {\dd}z\,{\dd}z^*\,e^{z^*
z}\,\bigl\l z^* |\,e^S\, e^{-\mu H^\pm}\, e^T \,e^{-\nu H^\pm} | z
\bigr\r\,. \label{bm:eq71}
\end{equation}
For the sake of simplicity we shall consider the $XX$ model only
and take those $H^\pm$ that correspond to $H$ (\ref{bm:eq1}) at
$\ga=0$.

To represent R.H.S. of (\ref{bm:eq71}) as the path integral, we
first introduce new coherent states $|x(I)\big\r$, $\big\l x^*(I)
|$, where $2 L\times M$ independent Grassmann parameters are
arranged in the form of $2 L$ ``vectors'' $x^*(I)$, $x(I)$
($I\in\{1, \dots, L\}$). It allows to insert $L$ times the
decompositions of unity
\[
\int{\dd}x^*(I){\dd}x(I)\,\exp\bigl(-x^*(I)x(I)\bigr)\,|x(I)\big\r
\big\l x^*(I) |
\]
into R.H.S. of (\ref{bm:eq71}). We define then the additional
variables satisfying the quasi-periodicity conditions:
\begin{equation}
-{\h E}\,x(0)\,=\,x(L+1)\equiv z\,,\qquad
-\,x^*(L+1)\,=\,x^*(0)\,{\h E}^{-1}\,\equiv z^*\,. \label{bm:eq8}
\end{equation}
Here, ${\h E}\,\equiv\,e^{\h S}\,e^{-\mu {\h H}^\pm}\,e^{\h T}$ with
the matrices ${\h H}^\pm$ expressed \cite{m3} through the hopping
matrices (\ref{bm:eq2}): ${\h H}^\pm = - {\h\Dl}^{(\mp)} + h {\h
I}$, where ${\h I}$ is a unit $M\times M$ matrix. The described
procedure allows to pass in the limit $L\to\infty$ from $(L+1)$-fold
integration to the {\it continuous} one over ``infinite'' product of
the measures ${\dd}x^*(\tau){\dd}x(\tau)$ on a space of trajectories
$x^*(\tau)$, $x(\tau)$, where $\tau\in {\mathbb R}$:
\[
{\cG}^\pm_{\F}\,Z^\pm_{\F}\,=\,\int\,e^S\,{\dd}\la^*\,{\dd}\la
\prod\limits_{\tau}{\dd}x^*(\tau){\dd}x(\tau)\,.
\]
The integration over the auxiliary Grassmann variables $\la^*$,
$\la$ guarantees the fulfilment of the continuous version of the
constraints (\ref{bm:eq8}). The action functional is
$S\,\equiv\,\int L(\tau)\,d\tau$, where $L(\tau)$ is the Lagrangian:
\[
\begin{array}{l} \displaystyle{L(\tau)\,\equiv\,
x^*(\tau)\,\Bigl(\frac{{\dd}}{{\dd}\tau}\,-\,{\h
H}^\pm\Bigr)\,x(\tau)
\,+\,J^*(\tau) x(\tau)\,+\,x^*(\tau) J(\tau)}\,,\\[0.3cm]
J^*(\tau)\,\equiv\,\la^* \bigl(\dl(\tau)\,{\h I}\,+\,
\dl(\tau-\nu)\,{\h E}^{-1}\bigr)\,,\quad
J(\tau)\,\equiv\,\bigl(\dl(\tau)\,{\h I}\,+\, \dl(\tau-\nu)\,{\h
E}\bigr)\,\la \,.
\end{array}
\]
The $\dl$-functions reduce $\tau$ to the segment $[0, \be]$. The
stationary phase requirements $\dl S/\dl x^*=0$, $\dl S/\dl x=0$
yield the regularized answer \cite{m3}:
\[
{\cG}^\pm_{\F}\,=\,{\det}\Bigl({\h I}\,+\,\frac{ e^{(\be-\nu){\h
H}^\pm}\,e^{\h S}\,e^{-\mu {\h H}^\pm}\,e^{\h T}\,-\,{\h I}}{{\h
I}\,+\,e^{\be\,{\h H}^\pm}} \Bigr)\,,
\]
The remainder correlators $G_{a b}(m, t)$ (\ref{bm:eq3}) (with $a,
b \in\{+, -\}$) are obtained analogously.

\section{Random walks}\label{bm:sec4}

The evolution of the states obtained by selective flipping of the
spins governed by the $XX$ Hamiltonian $H_0$ (\ref{bm:eq1}) is
related to a model of a random turns vicious walkers \cite{bog,
bog1}. Indeed, let us consider the following average over the
ferromagnetic state vectors $\langle \Uparrow |$, $| \Uparrow
\rangle$:
\begin{equation}
F_{j;\,l}(\la)\equiv \langle \Uparrow |\,\sigma _{j}^{+}e^{-\la
H_0} \sigma_{l}^{-} | \Uparrow \rangle \,,\label{bm:eq9}
\end{equation}
where ${|\Uparrow \rangle\equiv\otimes_{n=1}^M | \uparrow \rangle
_n }$, i.e., all spins are up, and $\la$ is an ``evolution''
parameter. Spin up (or down) corresponds to empty (or occupied)
site. Differentiating ${F_{j;\,l}({\la})}$ (\ref{bm:eq9}) and
applying the commutator ${\lbrack H_0,\sigma_j^{+} \rbrack}$, we
obtain the differential-difference equation ({\it master
equation}):
\begin{equation}
\displaystyle{ \frac{{\dd}}{{\dd}\la}\,F_{j;\,l}({\la})}\,=\,
{\displaystyle \frac12\,\bigl(
F_{j+1;\,l}({\la})\,+\,F_{j-1;\,l}({\la})\bigr)}
\,.\label{bm:eq10}
\end{equation}
The average $F_{j;\,l}$ may be considered as the generating
function of paths made by a random walker travelling from $l^{\rm
th}$ to $j^{\rm th}$ site. Really, its $K$-th derivative has the
form
\[
\displaystyle{\frac{{\dd}^K}{{\dd} {\la}^K} \,F_{j;\,l}({\la}) }\,
{\Biggl |}_{{\la} =0}\!\!\!=\langle \Uparrow \mid
\sigma_{j}^{+}(-\,H_0)^K\sigma_{l}^{-}\mid \Uparrow \rangle
=\!\!\!\! {\displaystyle \sum\limits_{n_1,\dots , n_{K-1}}}
{\Dl}^{(+)}_{j n_{K-1}}\dots {\Dl}^{(+)}_{n_2 n_1} {\Dl}^{(+)}_{n_1
l}\,.
\]
A single step to one of the nearest sites is prescribed by the
hopping matrix (\ref{bm:eq2}) with $s=+$. After $K$ steps, each path
connecting $l^{\rm th}$ and $j^{\rm th}$ sites contributes into the
sum. The $N$-point correlation function ($N\le M$),
\begin{equation}
F_{j_1, j_2,\dots, j_N;\, l_1, l_2,\dots, l_N}({\la})\,=\,\langle
\Uparrow \mid \sigma_{j_1}^{+} \sigma_{j_2}^{+}\dots
\sigma_{j_N}^{+}\,e^{-\la H_0}\,\sigma_{l_1}^{-}
\sigma_{l_2}^{-}\dots \sigma_{l_N}^{-}\mid \Uparrow \rangle
\,,\label{bm:eq11}
\end{equation}
enumerates the nests of the lattice paths of $N$ random turns
vicious walkers being initially located at the positions $l_1>
l_2>\dots > l_N$ and, eventually, at $j_1> j_2>\dots > j_N$. It is
expressed in the form \cite{bog}:
\begin{equation}
F_{j_1, \dots, j_N;\,l_1, \dots, l_N} ({\la}) = {\det} \bigl(
F_{j_r;\,l_s}({\la})\bigr)_{1\le r,\,s\le N}\,.\label{bm:eq12}
\end{equation}

The ground and the excited states of the $XX$ chain at  $h=0$ with
the total spin equal to $(M/2)-N$  are decomposed over a basis of
states $\sigma_{l_1}^{-} \sigma_{l_2}^{-}\dots \sigma_{l_N}^{-} \mid
\Uparrow \rangle$ with $N$ spins flipped \cite{iz}. Therefore, the
trace ${\widetilde F}_{m+1;\,1}({\la})\,\equiv\,\Tr
\bigl(\sigma_{m+1}^{+}e^{-{\la} H_0}\sigma_1^{-} \bigr)$ is a linear
combination of the generating functions (\ref{bm:eq11}) describing
the evolution of $N+1$ random turns walkers. The initial and the
final positions of one of them are fixed at $l_1=1$ and $j_1=m+1$,
respectively, while for the rest ones these positions are random. In
the thermodynamic limit, the number of the {\it virtual} walkers
tends to infinity. We apply the procedure described  in
\ref{bm:sec3} to calculation of ${\widetilde F}_{m+1;\,1}({\la})$
in the limit when $M$ and $N$ are large enough. In this limit, the
contribution of the terms with the subindex `B' become, with regard
at (\ref{bm:eq7}), negligible in (\ref{bm:eq6}). We thus obtain:
\[
\begin{array}{rcl}
 {\widetilde F}_{m+1;\,1}({\la})&=&\displaystyle{\Bigl[\tr
\bigl(e^{-{\la} {\h H}^{0}} {\widehat e}_{1,
m+1}\bigr)-\,\frac{\dd}{\dd {\al}}\Bigr]\,{\det}\Bigl({\h I}
+{\h\CU}_m+\frac{\al}{M}{\h\CV}_{m}\Bigr) \Bigg|_{{\al} =
0} }\\[0.5cm]
&=& \displaystyle{{\det}\bigl({\h I} + {\h\CU}_m \bigr)\left[\tr
\bigl(e^{-{\la} {\h H}^{0}} {\h e}_{1,
m+1}\bigr)-\frac{1}{M}\,\tr\Bigl(\frac{{\h\CV}_{m}}{{\h I} +
{\h\CU}_m}\Bigr)\right]}\,,
\end{array}
\]
where ${\h e}_{1, m+1}$ $\equiv$ $({\dl}_{1, n} {\dl}_{m+1,
l})_{1\le n, l\le M}$, and the matrix ${\h H}^{0}$ is used instead
of ${\h H}^{\pm}$ since $s$ can be taken zero at large enough $M$.
The traces of $\la$-dependent $M\times M$ matrices ${\h\CU}_m$ and
${\h\CV}_m$ are given below. Differential equation analogous to
(\ref{bm:eq10}) is fulfilled by ${\widetilde F}_{m+1;\,1}({\la})$.
At large separation $m$ it takes the form:
\begin{equation}
\displaystyle{\frac{\dd}{{\dd} {\la}}{\widetilde
F}_{m+1;\,1}({\la})} = {\displaystyle\frac12 \bigl({\widetilde
F}_{m;\,1}({\la}) + {\widetilde F}_{m+2;\,1}({\la})}\bigr) -
\Tr\bigl( H_0\,\sigma_{m+1}^{+}\,e^{-{\la} H_0}\,\sigma_{1}^{-}
\bigr)\,. \label{bm:eq13}
\end{equation}
We expand formally ${\widetilde F}_{m+1;\,1}({\la})$ with respect to
${\h\CU}_m$ and obtain the answer in two lowest orders as follows:
\begin{equation}
\begin{array}{rcl}
{\widetilde F}_{m+1;\,1}({\la}) &\approx& F_{m+1;\,1}({\la})\,+\,
F_{m+1;\,1}({\la})\times\tr{\h\CU}_m\,
-\,\displaystyle{\frac{1}{M}\,\tr{\h\CV}_{m}}\,,\\[0.2cm]
\tr{\h\CU}_m &=&{\displaystyle
(M\,-\,2 m)\,F_{1;\,1}({\la})}\,,\\[0.1cm]
\displaystyle{\frac{1}{M}}\,\tr{\h\CV}_{m}&=&{\displaystyle
F_{m+1;\,1}(2 {\la})\,-\,2\sum\limits_{l=1}^{m}
F_{m+1;\,l}({\la})\,F_{l;\,1}({\la})}\,.
\end{array}
\label{bm:eq14}
\end{equation}
Although $M$ and $m$ are chosen to be large in this expansion, the
ratio $m/M$ is kept bounded. In each order the master equation
(\ref{bm:eq13}) is fulfilled by (\ref{bm:eq14}). The contribution of
the second order can be re-expressed through the two-point functions
${F}_{m+1,l;\,l, 1}({\la})$ (see (\ref{bm:eq11}), (\ref{bm:eq12})).
Thus, summation over intermediate positions (of a virtual walker
located at $l^{\rm th}$ site) arises in the second order. A similar
picture is expected in next orders.

\section{Acknowledgement}\label{bm:sec5}
{\small One of us (C. M.) is grateful to the Organizers of 9th
International Conference ``Path Integrals -- New Trends and
Perspectives''. This work was partially supported by the RFBR, No.
07--01--00358.

\end{document}